\documentclass{basi}
\usepackage[british]{babel}
\usepackage[varg]{txfonts}
\usepackage{alltt}
\usepackage{rotating}
\usepackage{dcolumn}
\usepackage{longtable}
\usepackage{graphicx}
\usepackage{psfig}

\newcommand{\lapp}{\mbox{\raisebox{-0.3em}{$\stackrel{\textstyle <}{\sim}$}}}

\begin{document}
\setcounter{page}{1}

\title[Double-double radio galaxies from the FIRST survey]
      {Double-double radio galaxies from the FIRST survey}
\author[S.~Nandi and D.J. Saikia]%
       {S.~Nandi$^{1,2}$\thanks{email: \texttt{sumana@aries.res.in (SN)}}
         and D.~J.~Saikia$^{3,4}$\thanks{email: \texttt{djs@ncra.tifr.res.in (DJS)}} \\
       $^1$Aryabhatta Research Institute of Observational Sciences (ARIES), Manora Peak, Nainital, 263 129, India\\
       $^2$Department of Physics, Kumaun University, Nainital 263 001, India\\
       $^3$National Centre for Radio Astrophysics, TIFR, Pune University Campus, Post Bag 3, Pune 411 007, India\\
       $^4$Cotton College State University, Panbazar, Guwahati 781 001, India}

\pubyear{2012}
\volume{40}
\status{ in press}

\date{Received 2012 May 25; accepted 2012 July 21}

\maketitle
\label{firstpage}

\begin{abstract}
The radio structures and optical identifications of a sample of 242 sources
classified as double-double radio sources by Proctor (2011) from a morphological
study of sources in the FIRST (Faint Images of the Radio Sky at Twenty centimeters) 
survey (2003 April release, 811,117 entries) have been examined. We have been
able to confirm only 23 of these as likely to be double-double radio galaxies (DDRGs),
whose structures could be attributed to episodic nuclear activity in their host galaxies.
A further 63 require either higher-resolution radio observations or optical identifications 
to determine whether these are DDRGs. The remaining sources are unlikely to
be DDRGs. We have examined the luminosities, sizes and symmetry parameters of the
DDRGs and the constraints these place on our understanding of these sources.
\end{abstract}


\section{Introduction}\label{sec:Intro}
An important and interesting question in our understanding of active galactic nuclei (AGN) is
whether their nuclear activity is usually episodic in nature. There have been suggestions that
black holes grow during AGN phases with the total life time of the active phases ranging from
$\sim1.5\times10^8$ to $10^9$ yr (e.g. Marconi et al. 2004). Recurrent activity could also have significant
implications in feedback processes in active galaxies and on the evolution of galaxies themselves.
Although there has been increasing evidence of recurrent activity in AGN from both radio and
X-ray observations (see Saikia \& Jamrozy 2009 for a review), it is not clear how ubiquitous
this phenomenon might be. For example from deep low-frequency observations Sirothia et al.
(2009) did not find any unambiguous evidence of recurrent activity in a sample of 374 small-
sized sources. Marecki (2012) has interpreted the structure of the highly asymmetric giant 
radio galaxy J1211+743 (Pirya et al. 2011) to be due to recurrent nuclear activity.
To understand these varied aspects as well as the range of time scales of episodic
activity and possible reasons for it (see Kaiser, Schoenmakers \& R\"ottgering 2000; Czerny et al.
2009; Brocksopp et al. 2011), it is necessary to significantly increase the number of sources with
evidence of recurrent activity beyond the couple of dozen or so that are presently known (see
Saikia \& Jamrozy 2009). At present, there appears to be a reasonably wide range of time scales
of episodic activity which has been estimated from spectral and dynamical ages of the outer and
inner lobes of a few DDRGs. These range from $\sim10^5$ yr for 3C293 (Joshi et al. 2011) to $\sim10^8$ yr
for the largest Mpc-scale DDRGs (e.g. Schoenmakers et al. 2000; Konar et al. 2006).

Approximately 10 per cent of AGN are luminous at radio wavelengths, referred to as radio-
loud objects. Drawing an analogy with X-ray binary systems in our Galaxy, a number of authors
have suggested that radio activity in AGN could itself be episodic in nature (Nipoti, Blundell \&
Binney 2005; K\"ording, Jester \& Fender 2006). One of the clearest signs of episodic activity is
seen in radio-loud objects where there are more than one pair of distinct outer lobes, which can
be unambiguously ascribed to different cycles of activity. Although most of these sources exhibit
two cycles of activity, and are referred to as DDRGs, there are two
examples of sources which appear to possibly exhibit three cycles of activity, namely B0925$+$420
(Brocksopp et al. 2007) and J140948.85$-$030232.5 (Hota et al. 2011), the latter being associated
with a spiral host galaxy. Almost all the sources exhibiting evidence of a double-double structure
are associated with galaxies, with the possible exception of 4C02.27 which is associated with a
quasar (Jamrozy, Saikia \& Konar 2009). Evidence of episodic activity has also been reported
from a combination of X-ray and radio observations, where inverse-Compton scattered X-ray
emission from old electrons in the relic lobes are seen along with synchrotron emission from the
recent cycle of activity. Sources where this has been reported include 3C191 and 3C294 (Erlund
et al. 2006), as well as the well-studied object Cygnus A (Steenbrugge, Blundell \& Duffy 2008;
Steenbrugge, Heywood \& Blundell 2010).

To understand the nature of these sources and possible reasons for their episodic activity, 
we need to enlarge the sample of objects. As a first step we have focused on the FIRST survey
(Becker, White \& Helfand 1995), where Proctor
(2011) has done a classification of the structures of sources into different categories. Of interest
here are the 242 sources Proctor (2011) has classified as DDRGs based on the identification of at
least four different components in the radio images. Since the mere existence of four components
does not guarantee the source to be a DDRG, we have examined the radio structure as well as
the optical identifications of all the 242 sources to identify those we believe to be good 
examples of DDRGs, possible examples which require further observations and also sources 
which do not appear to be DDRGs. We describe briefly the methodology we have adopted in 
Section 2, describe the DDRGs identified from this survey in Section 3, and discuss the 
nature of these sources in Section 4.

\section{Methodology of classification of the DDRGs}

As mentioned earlier, we have examined the radio structures and optical identifications
of each of the 242 sources listed by Proctor (2011) as DDRGs. To identify the optical objects
we have used the Sloan Digital Sky Survey (SDSS) Data Release 8 (DR8)\footnote {The SDSS Web Site is 
http://www.sdss.org/}. If the source is not covered 
in DR8 we have examined the Digital Sky Survey DSS R-band images to find the optical identifications. 
We consider the optical object to be identified with the radio core component if the optical 
position is within an arcsec of the radio peak position of an unresolved or slightly resolved compact 
component. Normally the difference is within $\sim$0.5 arcsec. However, in some cases the presence of 
extended emission near the core at 1400 MHz may shift the centroid and higher frequency 
observations are required to locate the core more accurately.
In sources without a radio core we consider the optical object to be identified with 
the radio source if it lies close to the axis defined by the inner lobes. The extended lobe-like 
emission in the source J1158$+$2625 is not visible in the NRAO VLA Sky Survey (NVSS) image and 
is likely to be spurious. This source has not been considered further.

Sources where radio emission from two cycles of activity are clearly distinguishable
with the inner structures being more compact, and with an optical identification
located between the inner doubles have been classified as DDRGs. Such sources
without an optical identification have been classified as candidate DDRGs, as illustrated
in the  Appendix.
In these cases deep optical imaging is required to examine whether one of
the inner compact components may be co-incident with
an optical identification. Deep multi-frequency radio observations would also help determine
whether the radio spectrum is flat, as is usually the case for radio cores. Alternatively it could
help detect possible backflow in the form of tails if these compact inner components are hot-spots.
 
A source where the optical object is co-incident with one of the inner compact
components is clearly not a DDRG, and is listed as a non-DDRG in this study. 
For example the sources J0759$+$4051 and J1434$+$0441 discussed in the Appendix
belong to this category.

A DDRG is expected to have diffuse outer lobes due to the an earlier cycle of
activity and more compact hot-spots in the inner lobes which are due to 
a more recent cycle of AGN activity. With inadequate resolution the 
components of radio emission on either side of the host galaxy may not be
resolved clearly to reveal the detailed structure. These components
may be due to either two cycles of activity or the secondary inner peaks
may be caused by backflow from the outer hot-spot. We have fitted two dimensional
gaussians to all such sources. Those where the inner components are clearly 
resolved and more extended than the outer ones are likely to be due to peaks of
emission caused by back-flowing plasma and have been classified as non-DDRGs; 
those where the inner components are more compact but higher resolution observations 
are required to determine their structure more reliably are referred to as 
candidate DDRGs in our study. J0032$-$0019, shown in the Appendix, represents the kind 
of sources where the inner components are due to backflowing plasma from the hot-spots.

The wide-angle tailed sources (WATs) often have more compact emission closer to the host galaxy
with more diffuse tails of emission extending from these features (e.g. Blanton et al. 2003; Mao et
al. 2010). Detailed spectral studies are required to establish the episodic nature of these sources.
These WATs are listed separately in this paper and will not be discussed here.

\begin{table}
\caption{DDRGs from the FIRST survey}
\label{tab:sample_ddrgs}
{\small
\begin{tabular}{c c c c c c c c }
\hline
Source     & Opt. &  $z$    &  RA$_{\rm optical}$ & Dec$_{\rm optical}$ & RA$_{\rm core}$ & Dec$_{\rm core}$ &  Notes    \\
name       & Id.$^\oplus$  &         &  hh:mm:ss.ss        & dd:mm:ss.ss          &  hh:mm:ss.ss    &  dd:mm:ss.ss     &           \\
(1)        & (2)  &  (3)    &    (4)              &   (5)                  &  (6)            & (7)              &    (8)   \\
\hline
J0746$+$4526 & G  & (0.517) &  07:46:17.92        & $+$45:26:34.47         &                 &                    &          \\
J0804$+$5809 &    &         &  08:04:42.79        & $+$58:09:34.94         &                 &                    &          \\
J0855$+$4204 & G  & (0.279) &  08:55:49.15        & $+$42:04:20.12         &                 &                    &         \\
J0910$+$0345 & G  & (0.588) &  09:10:59.10        & $+$03:45:31.68         &                 &                    &  1       \\
J1039$+$0536 & G  & 0.0908  &  10:39:28.21        & $+$05:36:13.61         &                 &                    &          \\
J1103$+$0636 & G  & (0.449) &  11:03:13.29        & $+$06:36:16.00         &                 &                    &          \\
J1158$+$2621 & G  & 0.1120  &  11:58:20.13        & $+$26:21:12.08         & 11:58:20.12     &   $+$26:21:12.04   &   2       \\
J1208$+$0821 & G  & (0.600) &  12:08:56.78        & $+$08:21:38.57         & 12:08:56.78     &   $+$08:21:38.44   &          \\
J1238$+$1602 & S  &         &  12:38:21.20        & $+$16:02:41.43         &                 &                    &          \\
J1240$+$2122 & G  & (0.357) &  12:40:13.48        & $+$21:22:33.04         &                 &                    &          \\
J1326$+$1924 & G  & 0.1762  &  13:26:13.67        & $+$19:24:23.75         &                 &                    &          \\
J1328$+$2752 & G  & 0.0911  &  13:28:48.45        & $+$27:52:27.81         & 13:28:48.43     &   $+$27:52:27.55   &   3       \\
J1344$-$0030 & G  & (0.579) &  13:44:46.92        & $-$00:30:09.28         &                 &                    &          \\
J1407$+$5132 & G  & (0.324) &  14:07:18.49        & $+$51:32:04.88         &                 &                    &          \\
J1500$+$1542 & G  & (0.456) &  15:00:55.18        & $+$15:42:40.64         &                 &                    &          \\
J1521$+$5214 & G  & (0.537) &  15:21:05.90        & $+$52:14:40.15         &                 &                    &          \\
J1538$-$0242 & G  & (0.575) &  15:38:41.31        & $-$02:42:05.52         &                 &                    &          \\
J1545$+$5047 & G  & 0.4309  &  15:45:17.21        & $+$50:47:54.18         &                 &                    &          \\
J1605$+$0711 & G  & (0.268) &  16:05:13.74        & $+$07:11:52.56         &                 &                    &          \\
J1627$+$2906 & G  & (0.722) &  16:27:54.63        & $+$29:06:20.00         &                 &                    &          \\
J1649$+$4133 & S  &         &  16:49:28.32        & $+$41:33:41.58         &                 &                    &          \\
J1705$+$3940 & G  & (0.701) &  17:05:17.83        & $+$39:40:29.25         &                 &                    &          \\
J1706$+$4340 & S  &         &  17:06:25.43        & $+$43:40:40.41         &                 &                    &          \\
\hline
\end{tabular}

\footnotesize
1: separation of optical position from nearest radio peak, $\Delta_{\rm radio-opt}$ is $\sim$2.5 arcsec;
2: known DDRG (Owen \& Ledlow 1997; Saikia \& Jamrozy 2009); 
3: misaligned DDRG. In all the Tables the photometric redshifts, taken from SDSS, are enclosed 
within parentheses.\\
$^\oplus$ In the second Column G and S represent galaxy and star respectively, as classified in SDSS.
Spectroscopic observations are required to determine whether the stellar objects are quasars.
}
\end{table}

\begin{figure}
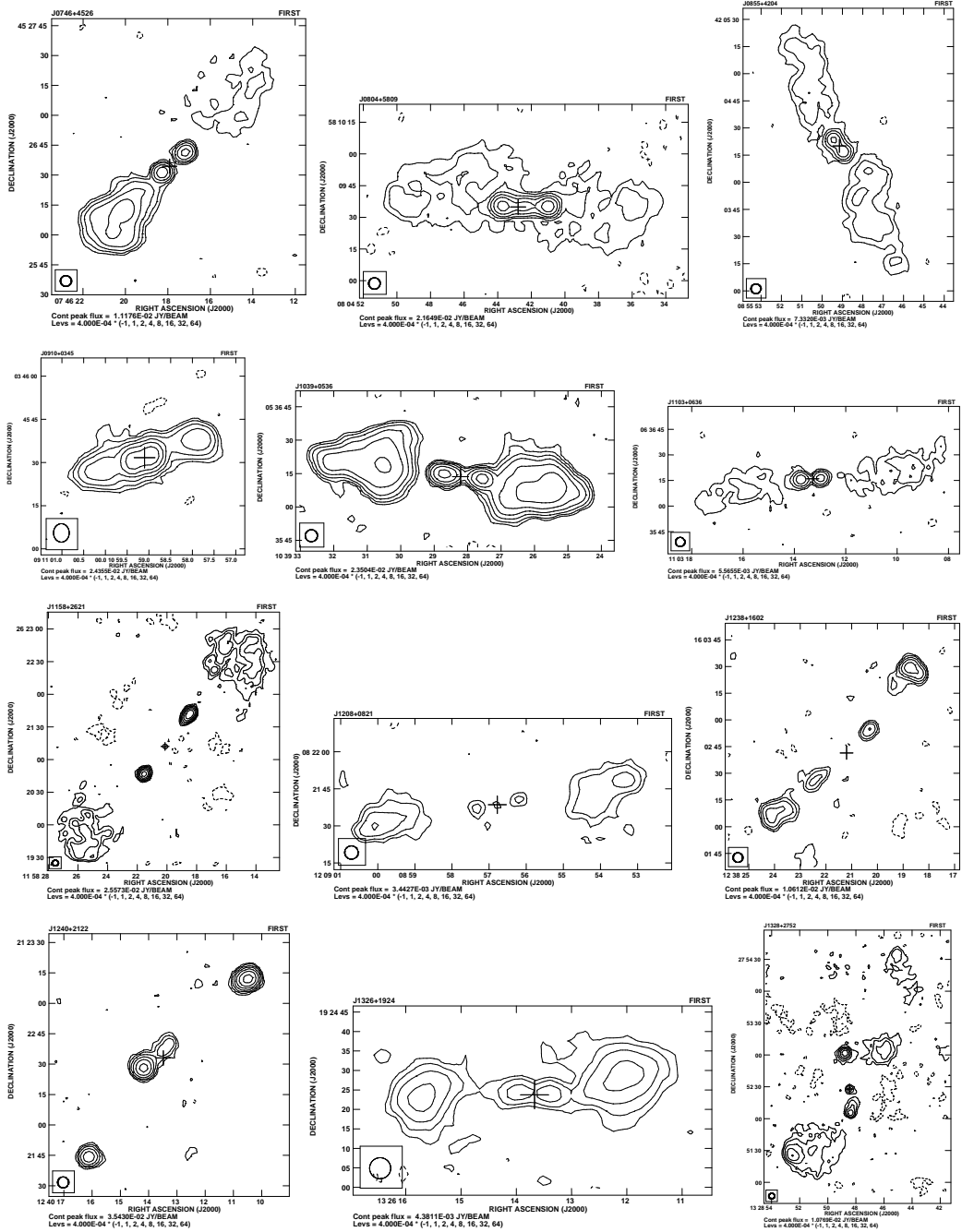

\vbox{
   \hbox{
      \psfig{file=OBJ13_DDRG1.PS,width=1.75in,angle=0}%
      \psfig{file=OBJ16_DDRG2.PS,height=1.4in,angle=-90}
      \psfig{file=OBJ19_DDRG3.PS,width=1.45in,angle=0}%
        }
\vspace{0.1cm}
   \hbox{
      \psfig{file=OBJ22_DDRG4.PS,width=1.4in,angle=-90}%
      \psfig{file=OBJ41_DDRG5.PS,height=1.2in,angle=-90}
      \psfig{file=OBJ49_DDRG7.PS,width=1.9in,angle=-90}%
 }
\vspace{0.1cm}
 \hbox{
      \psfig{file=OBJ63_DDRG9.PS,width=1.6in,angle=0}%
      \psfig{file=OBJ69_DDRG10.PS,width=2.2in,angle=-90}%
      \psfig{file=OBJ77_DDRG11.PS,height=1.7in,angle=0}
        }
\vspace{0.1cm}
 \hbox{
      \psfig{file=OBJ79_DDRG12.PS,height=1.8in,angle=0}
      \psfig{file=OBJ93_DDRG13.PS,height=1.4in,angle=-90}
     \psfig{file=OBJ95_DDRG14.PS,height=1.8in,angle=0}
        }

}
\caption[]{ FIRST images of the DDRGs listed in Table 1. In all the Figures the $+$ sign when shown denotes
the position of the optical object.
}
\end{figure}
\setcounter{figure}{0}
\begin{figure}
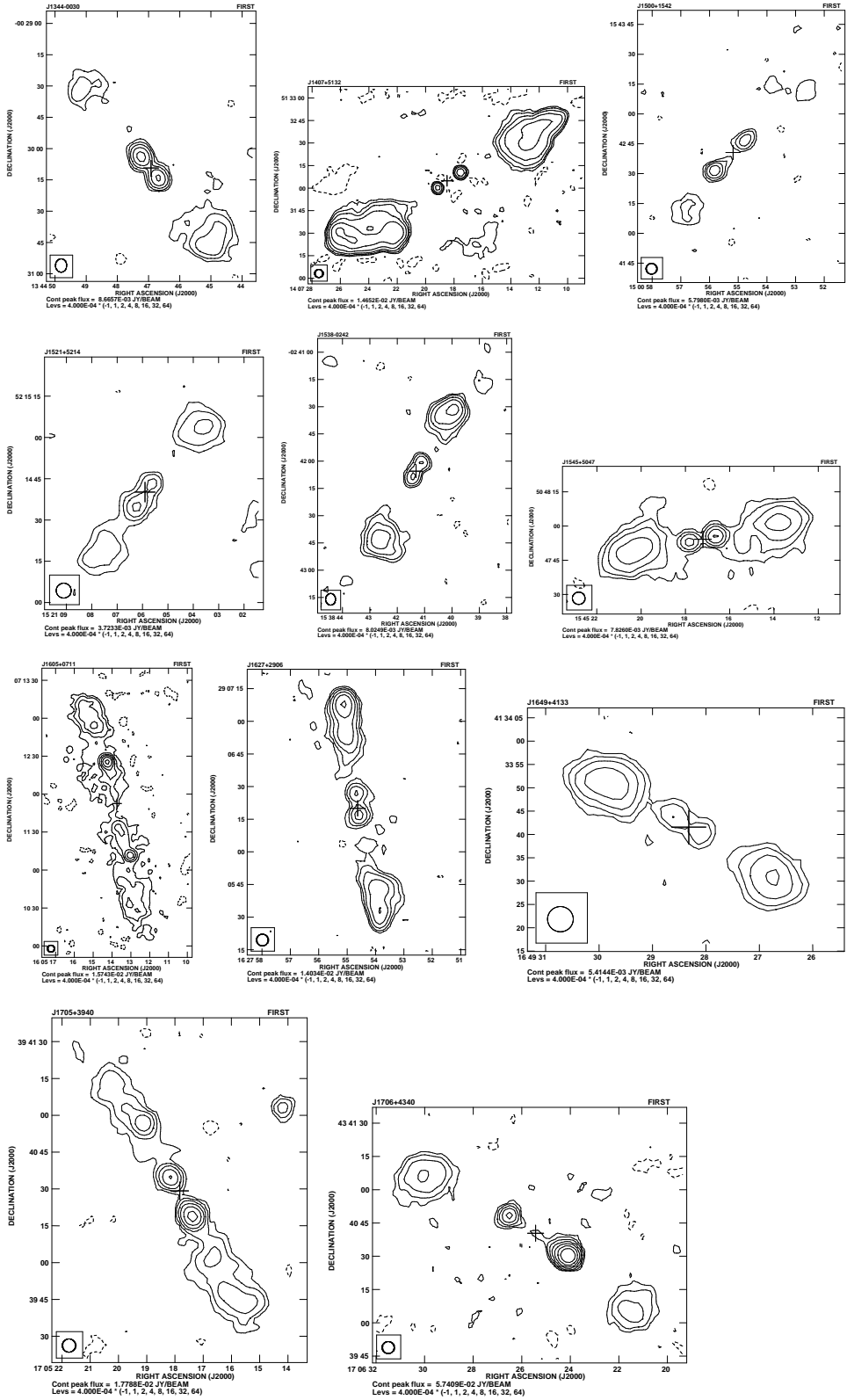

\vbox{

 \hbox{
       \psfig{file=OBJ109_15.PS,height=1.9in,angle=0}
      \psfig{file=OBJ120_DDRG16.PS,width=1.9in,angle=-90}
      \psfig{file=OBJ155_DDRG17.PS,height=1.9in,angle=0}
        }
\vspace{0.1cm}
 \hbox{
       \psfig{file=OBJ162_DDRG18.PS,height=1.8in,angle=0}
      \psfig{file=OBJ168_DDRG19.PS,height=1.9in,angle=0}
      \psfig{file=OBJ173_DDRG20.PS,width=1.9in,angle=-90}

     }
\vspace{0.1cm}
 \hbox{
      \psfig{file=OBJ186_DDRG21.PS,height=2.0in,angle=0}
      \psfig{file=OBJ202_DDRG22.PS,height=2.0in,angle=0}
     \psfig{file=OBJ212_DDRG23.PS,width=2.2in,angle=-90}

     }
\vspace{0.1cm}
 \hbox{
      \psfig{file=OBJ220_DDRG24.PS,height=2.4in,angle=0}
     \psfig{file=OBJ222_DDRG25.PS,width=2.2in,angle=-90}

}

}
\caption[]{(continued)}
\end{figure}

This leaves us with a sample of 23 DDRGs and 63 candidate DDRGs, where further observations 
are required to establish whether these might be DDRGs. The list of DDRGs are presented
in Table 1, some of its properties in Table 2, while the list of candidate DDRGS are presented 
in Table 3. The list of non-DDRGs
and the WAT sources are presented in the Appendix (Tables 4 and 5 respectively). 
These tables are arranged as follows. Column 1: source name; Column 2: optical identification; 
Column 3: redshift; Columns 4 and 5: The right ascension (hh:mm:ss.ss) and declination (dd:mm:ss.ss) 
of the optical objects 
in J2000 co-ordinates; Columns 6 and 7: The right ascension (hh:mm:ss.ss) and declination (dd:mm:ss.ss) of the radio
core positions; Column 8: Notes on individual sources. Since we are not discussing the WAT sources in detail, notes
are not provided for Table 5. 
 
Some of the observed properties of the FIRST DDRGs are listed in Table 2 which is arranged as follows. 
Column 1: source name; Column 2: optical identification; Column 3: redshift; Columns 4 and 5: 
projected linear size of the inner and outer double-lobed sources in kpc; Column 6: the locations 
of the components farther/closer from the core for the inner double. The symmetry parameters mentioned in Columns
7$-$10 are all in the same sense as given in Column 6. Columns 7 and 8 represent respectively 
the armlength or separation ratio for the 
inner and outer doubles; Columns 9 and 10: flux density ratios for the inner and outer doubles; and 
Columns 11 and 12: log of radio luminosity at an emitted frequency of 1400 MHz for the inner and outer doubles. 
The FIRST images of all the DDRGs are shown in Fig. 1.

\begin{table}
\caption{Some of the observed properties of the sample of DDRGs}
\label{tab:ddrgs}
\small
\centering
\begin{tabular}{l  c  l  rr  c rr rr rr  }
\hline
Source       & Opt.& Red- & $l_{in}$& $l_{o}$& Cmp.& $R$$_{\theta(in)}$ &$R$$_{\theta(o)}$ & $R$$_{s(in)}$ &$R$$_{s(o)}$
&  $P$$_{in}$ & $P$$_{o}$ \\
             & Id. & shift& kpc  & kpc   &     &       &       &        &        &  W/Hz & W/Hz   \\
    (1)      & (2) &  (3) &  (4) &  (5)  & (6) &  (7)  &  (8)  &  (9)   &  (10)  &  (11) & (12)   \\
\hline
J0746$+$4526 &  G  &(0.517)& 95   & 630   &N/S  &1.86   &1.35   & 0.79   &0.19         & 25.41& 26.17     \\
J0804$+$5809 &     &       &      &       &W/E  &2.00   &0.99   & 1.43   &0.81         &      &           \\
J0855$+$4204 &  G  &(0.279)& 40   & 516   &N/S  &1.17   &0.86   & 1.34   &0.87         & 24.64& 25.30     \\
J0910$+$0345 &  G  &(0.588)& 34   & 225   &E/W  &2.17   &0.73   & 0.84   &0.99         & 25.85& 25.66     \\
J1039$+$0536 &  G  &0.0908 & 29   & 163   &W/E  &1.25   &0.74   & 1.03   &1.29         & 24.05& 24.97     \\
J1103$+$0636 &  G  &(0.449)& 71   & 652   &E/W  &1.28   &0.93   & 1.51   &0.72         & 24.99& 25.56     \\
J1158$+$2621 &  G  &0.1120 & 143  & 503   &N/S  &1.70   &1.01   & 1.81   &1.31         & 24.33& 24.80     \\
J1208$+$0821 &  G  &(0.600)& 121  & 687   &W/E  &1.07   &1.04   & 0.93   &0.96         & 24.49& 25.76     \\
J1238$+$1602 &  S  &       &      &       &S/N  &1.23   &0.91   & 1.12   &0.79         &      &           \\
J1240$+$2122 &  G  &(0.357)& 68   & 584   &S/N  &2.00   &1.07   & 4.40   &0.72         & 25.10& 25.42     \\
J1326$+$1924 &  G  &0.1762 & 27   & 149   &E/W  &1.20   &1.25   & 1.06   &0.65         & 23.68& 24.56     \\
J1328$+$2752 &  G  &0.0911 & 95   & 355   &N/S  &1.53   &1.61   & 1.85   &0.63         & 23.76& 24.39     \\
J1344$-$0030 &  G  &(0.579)& 86   & 643   &N/S  &1.25   &1.06   & 1.30   &0.35         & 25.46& 25.60     \\
J1407$+$5132 &  G  &(0.324)& 87   & 613   &W/E  &1.40   &0.84   & 1.51   &0.62         & 24.43& 26.15     \\
J1500$+$1542 &  G  &(0.456)& 128  & 485   &S/N  &1.54   &0.76   & 1.68   &1.27         & 24.93& 24.92     \\
J1521$+$5214 &  G  &(0.537)& 64   & 391   &S/N  &1.50   &0.74   & 2.09   &0.81         & 24.99& 25.28     \\
J1538$-$0242 &  G  &(0.575)& 59   & 534   &N/S  &1.16   &0.93   & 0.73   &1.24         & 25.05& 26.30     \\
J1545$+$5047 &  G  &0.4309 & 66   & 483   &E/W  &1.12   &0.84   & 0.67   &1.23         & 24.95& 25.69     \\
J1605$+$0711 &  G  &(0.268)&311   & 576   &S/N  &1.30   &1.06   & 0.54   &0.69         & 25.05& 25.38     \\
J1627$+$2906 &  G  &(0.722)& 74   & 702   &N/S  &2.20   &0.98   & 0.80   &1.46         & 25.40& 26.30     \\
J1649$+$4133 &  S  &       &      &       &E/W  &1.25   &0.94   & 1.38   &1.66         &      &           \\
J1705$+$3940 &  G  &(0.701)& 130  & 528   &S/N  &1.70   &0.96   & 2.52   &1.27         & 25.84& 26.07     \\
J1706$+$4340 &  S  &       &      &       &S/N  &1.24   &0.98   & 8.67   &0.59         &                  \\
\hline
\end{tabular}
\end{table}

\begin{figure}
\centering
\includegraphics[scale=0.3,angle=-90]{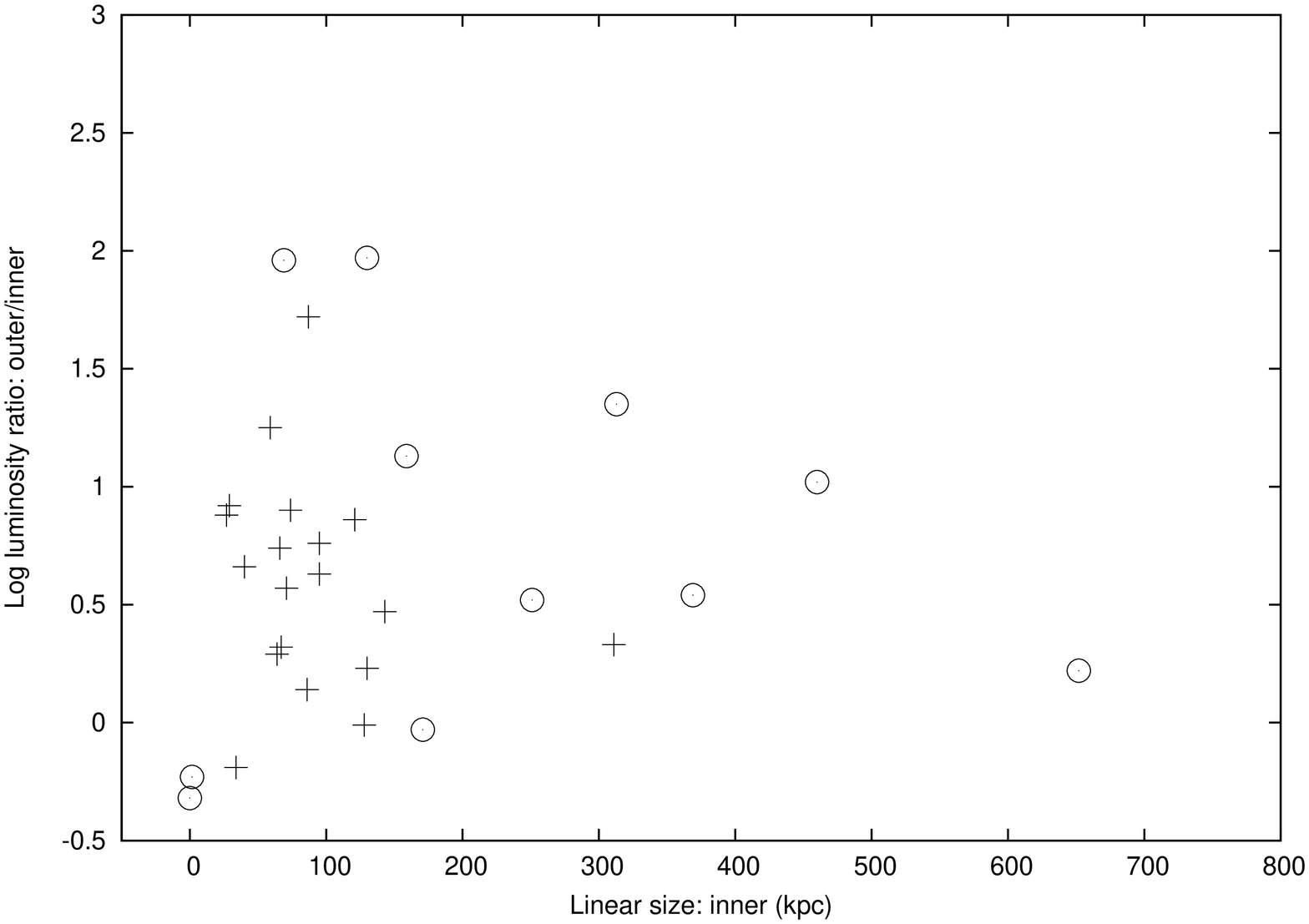}
\caption[]{The ratio of the luminosities of the outer to the inner doubles vs the projected linear size of the
inner double. The $+$ signs represent the DDRGs identified from the FIRST survey, while the open
circles represent those from Saikia et al. (2006).
}

\end{figure}

\begin{figure}
\centering
\includegraphics[scale=0.3,angle=-90]{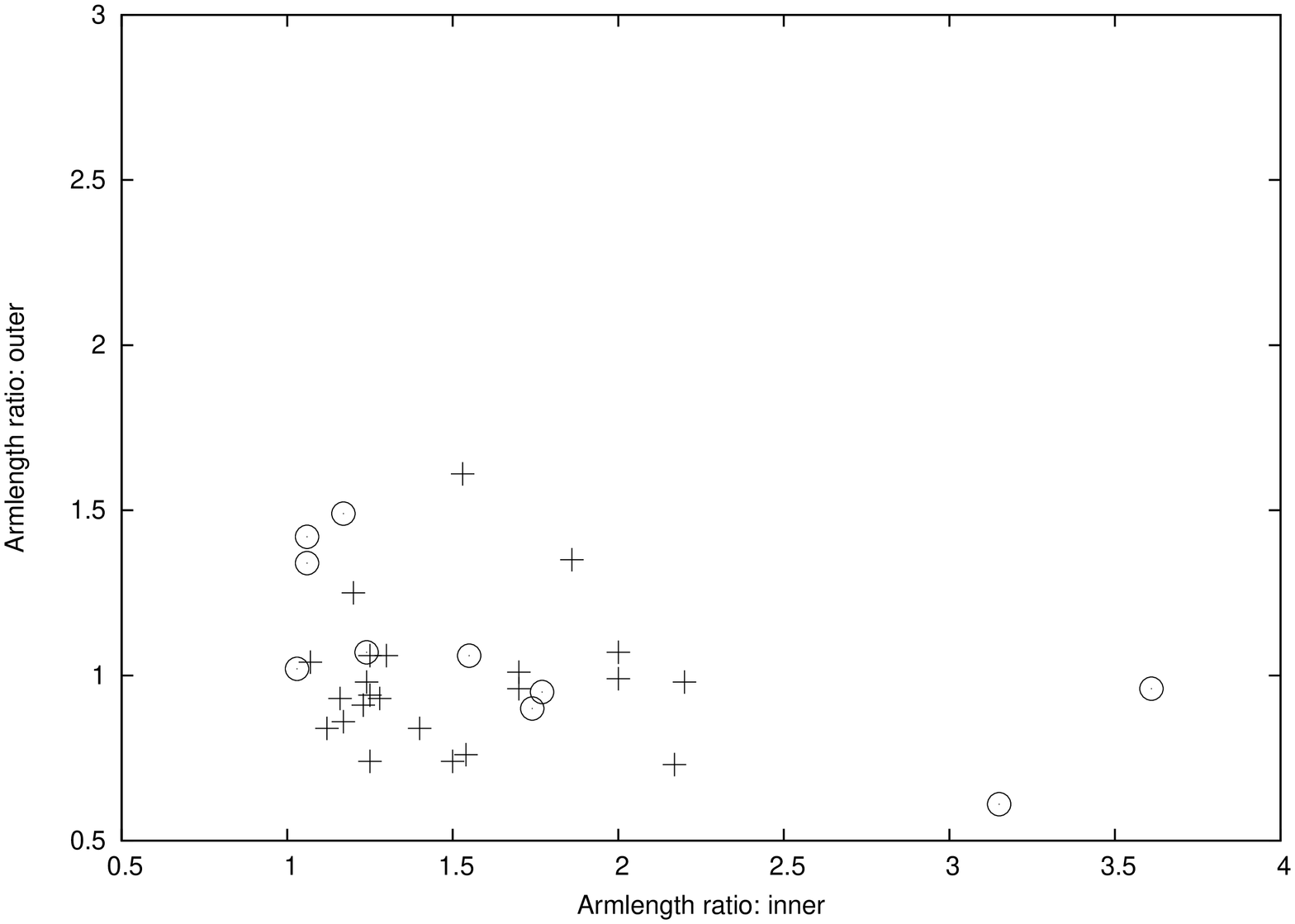}\\
\includegraphics[scale=0.3,angle=-90]{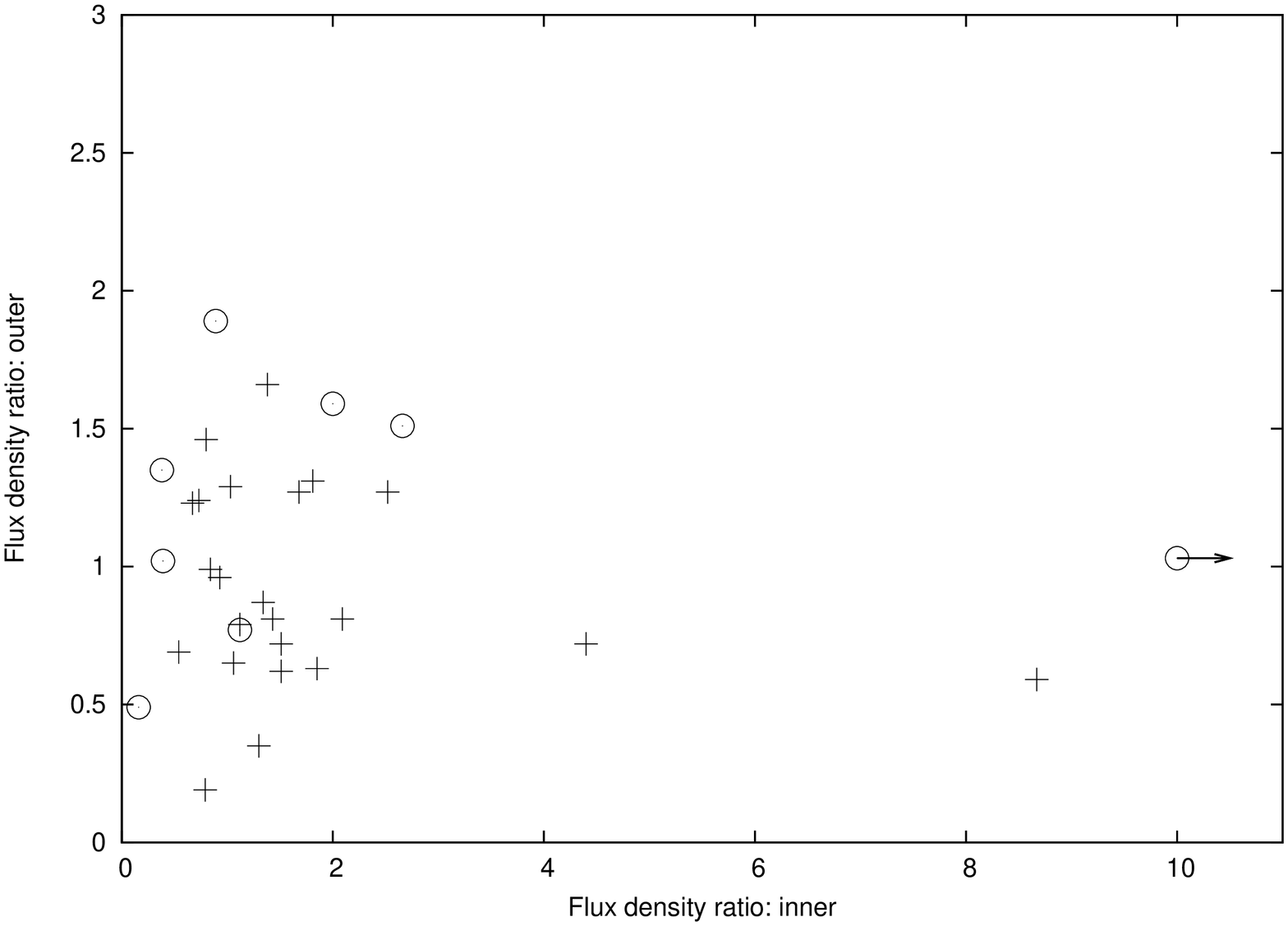}\\

\caption[]{Upper panel represents arm-length ratio of the outer double vs the arm-length ratio of the inner double where as 
Lower panel represents flux density ratio of outer double vs flux density ratio of inner double.
The $+$ signs represent the DDRGs identified from the FIRST survey, while the open circles represent
those from Saikia et al. (2006).
 }

\end{figure}

\section{Results and discussions}
\label{sec:res_and_dis}
\subsection{Linear sizes}
A comparison of the sizes of the DDRGs selected from the FIRST survey (Table 1) with 
the sample compiled by Saikia, Konar \& Kulkarni (2006) shows that the median sizes
of both the inner and outer doubles are smaller for the FIRST sources. Most of the
sources in the Saikia et al. (2006) compilation were from lower-resolution surveys
such as the NVSS, and there was a natural bias against selecting DDRGs with small
inner and outer doubles. The median values of the inner and outer doubles in the
FIRST DDRGs are $\sim$75 and 530 kpc respectively, compared with $\sim$165 and 1100
kpc from the Saikia et al. compilation. Studies of specific smaller sources 
such as 3C293 (Joshi et al. 2011) and 4C02.27 (Jamrozy, Saikia \& Konar 2009) 
suggest that time scales of episodic activity 
could be as small as $\sim$10$^5$ yr, compared with time scales of $\sim$10$^8$ yr
for Mpc-scale objects (e.g. Schoenmakers et al. 2000; Konar et al. 2006), making
it difficult to understand episodic activity in terms of periodic supply of fuel
by an interacting companion. The smaller sizes of FIRST DDRGs demonstrate clearly
that a wide range of time scales of episodic activity are possible, which may 
extend to even CSS and GPS scale objects as has been speculated for the 
archetypal GPS object CTA21 (Salter et al. 2010). 

In Fig. 2 we plot the ratio of the luminosities of the outer double to that of
the inner one versus the projected linear size of the inner double. Schoenmakers
et al. (2000) suggested an inverse correlation between these two parameters with
the outer lobes being more luminous than the inner ones,  while Saikia et al. (2006) 
found the inner ones to be more luminous in the smallest inner doubles and
the overall correlation to have a reduced level of significance. Considering the
FIRST DDRGs along with the Saikia et al. sample, we find that while there may be
an upper envelope to this diagram suggesting an inverse relation, the correlation
is not statistically significant. The range of the ratio of luminosities appears
large for sources $\lapp$200 kpc, varying by a factor of over 100. Note that the only
source in the new sample where the inner double is much more luminous than the
outer one has a size of only 34 kpc, the second smallest in the sample. 
The evolution of the smallest sources could be affected by the dense interstellar
medium of the host galaxy leading to more luminous components due to a more
efficient dissipation of energy as well as better confinement by the medium
(e.g. Jeyakumar et al. 2005, and references therein).  
   
\subsection{Arm-length and brightness ratios}
\label{sec:res_and_dis:compt_bright_asymmtery}
We have re-examined the trend for the inner doubles to be more asymmetric in both
brightness and location compared with the outer ones for the sample compiled by
Saikia et al. Such studies might provide clues towards understanding the 
environments in which these sources are evolving as well as any possible 
asymmetries in the oppositely-directed radio jets. For the FIRST sample the 
estimates of arm-length ratios have been made from the positions of the optical 
objects and it would be useful to make these estimates from higher frequency images 
with higher resolution where the cores are likely to be identified. Considering 
both the samples together there is a marginal trend for the inner double
to be more asymmetric in the location of the outer lobes, but the two 
distributions are not signficantly different (Fig. 3). However, the asymmetries in the outer and
inner lobes are often not in the same sense, suggesting that these are not likely to be due to the
effects of orientation and relativistic motion.

A comparison of the flux density ratios of the inner and outer doubles also shows this to
be not significantly different. There is also no significant trend for the farther component to be
brighter for either the inner or outer doubles. A closer brighter component is expected when the
jet on this side is interacting with denser material as it propagates outwards (Fig. 3).

\section{Concluding remarks}
\label{sec:con_remk}
Early studies of DDRGs suggested that these are likely to be associated with giant radio sources,
yielding time scales of episodic activity of $\sim$10$^8$ yr or so (e.g. Schoenmakers et al. 2001; Saikia,
Konar \& Kulkarni 2006; Saikia \& Jamrozy 2009 and references therein). Identification of 23
DDRGs from the FIRST survey has shown that these often occur in sources with overall projected 
sizes of hundreds of kpc as is seen in the misaligned DDRG 3C293 (Joshi et al. 2011).
Salter et al. (2010) also speculated that radio emission from the archetypal GPS source CTA21
may be episodic. Examples of DDRGs appear to occur over a wide range of size scales, and
surveys of different resolutions are required to be able to identify these objects. Detailed spectral 
and dynamical age studies will help us explore the range of time scales of episodic nuclear
activity. In this paper we also list the candidate DDRGs which require further observations to determine
whether it is a DDRG, and also the ones we classify as non-DDRGs and WATs in the Appendix.

\section*{Acknowledgments}
We thank Deanne Proctor, Marek Jamrozy, Dave Green, the reviewer for several helpful comments 
which have improved the paper significantly.  
SN thanks NCRA for hospitality while this work was done. The National Radio Astronomy
Observatory is a facility of the National Science Foundation operated under cooperative agreement 
by Associated Universities, Inc. Funding for the SDSS and SDSS-II has been provided
by the Alfred P. Sloan Foundation, the Participating Institutions, the National Science Foundation, 
the U.S. Department of Energy, the National Aeronautics and Space Administration, the
Japanese Monbukagakusho, the Max Planck Society, and the Higher Education Funding Council 
for England. The SDSS is managed by the Astrophysical Research Consortium for the Participating 
Institutions. The Participating Institutions are the American Museum of Natural History, 
Astrophysical Institute Potsdam, University of Basel, University of Cambridge, 
Case Western Reserve University, University of Chicago,
Drexel University, Fermilab, the Institute for Advanced Study, the Japan Participation Group,
Johns Hopkins University, the Joint Institute for Nuclear Astrophysics, the Kavli Institute for
Particle Astrophysics and Cosmology, the Korean Scientist Group, the Chinese Academy of Sciences 
(LAMOST), Los Alamos National Laboratory, the Max-Planck-Institute for Astronomy
(MPIA), the Max-Planck-Institute for Astrophysics (MPA), New Mexico State University, Ohio
State University, University of Pittsburgh, University of Portsmouth, Princeton University, the
United States Naval Observatory, and the University of Washington.

{}

\section*{Appendix}

\begin{figure}
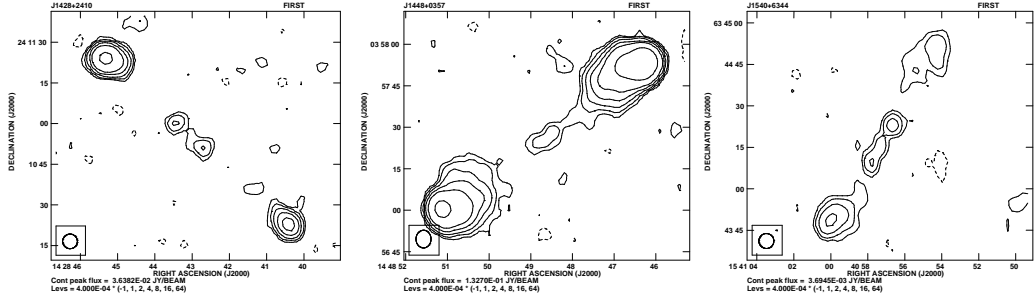

\vbox{
   \hbox{
      \psfig{file=OBJ129_CANDIDATEDDRG.PS,height=1.6in,angle=-90}
      \psfig{file=OBJ145_CANDIDATEDDRG.PS,height=1.6in,angle=-90}%
      \psfig{file=OBJ172_CANDIDATEDDRG.PS,height=1.6in,angle=-90}%
       }

}

\caption[]{ Examples of candidate DDRGs which require an optical identification to determine whether the
inner structure is due to either an inner double or a radio core and part/knot of a jet.
}


\end{figure}

\begin{figure}
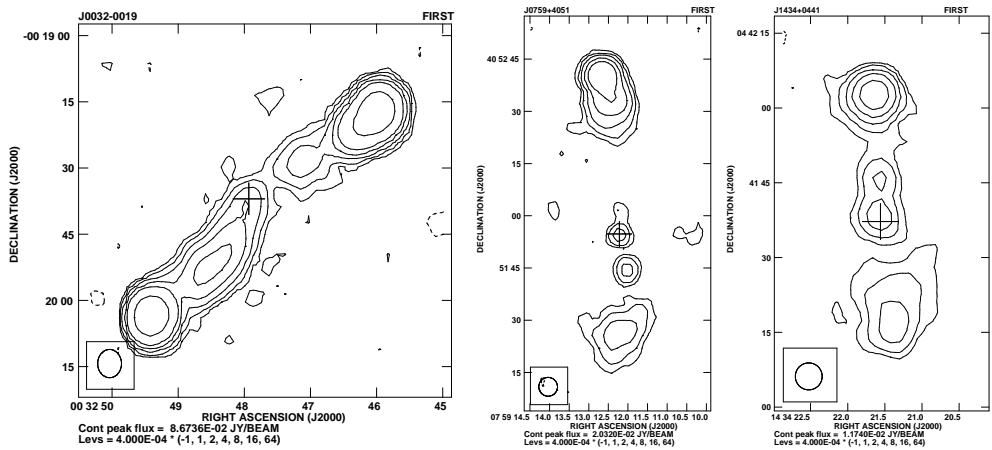

\vbox{
   \hbox{
      \psfig{file=OBJ3_NONDDRG.PS,height=2.4in,angle=-90}
      \psfig{file=OBJ15_NONDDRG.PS,height=2.4in,angle=0}%
      \psfig{file=OBJ137_NONDDRG.PS,height=2.4in,angle=0}%
       }

}

\caption[]{ Examples of sources which we have classified as non-DDRGs. The inner emission in J0032$-$0019
is likely to be due to backflow from the hot-spots, while in the case of J0759+4051 and J1434+0441, the
optical identification is co-incident with one of the components of the inner double which is likely to be the
radio core. The other feature is possibly a knot in the jet.
}


\end{figure}
In the Appendix, we list the candidate DDRGs from the FIRST survey which require further
observations to determine whether these might be DDRGs, the ones we classify as non-DDRGs
from the list given by Proctor (2011), and the wide-angle tailed sources. Examples of sources
which we classify as candidate DDRGs or non-DDRGs are shown in Figs. 4 and 5 respectively.


{\small
\bgroup

\newpage
\begin{center}

\begin{longtable}{cccccccc}
\caption{Candidate DDRGs from the FIRST survey.  } \label{tab:sample_candidateddrgs} \\
\hline
Source     & Opt. &  $z$    &  RA$_{\rm optical}$ & Dec$_{\rm optical}$ & RA$_{\rm core}$ & Dec$_{\rm core}$ &  Notes    \\
name       & Id.$^\oplus$  &         &  hh:mm:ss.ss        & dd:mm:ss.ss          &  hh:mm:ss.ss    &  dd:mm:ss.ss     &           \\
(1)        & (2)  &  (3)    &    (4)              &   (5)               &  (6)            & (7)              &   (8)   \\
\hline
\endfirsthead
{{ \tablename\ \thetable{} --(continued)}} \\

  \hline
Source     & Opt. &  $z$    &  RA$_{\rm optical}$ & Dec$_{\rm optical}$ & RA$_{\rm core}$ & Dec$_{\rm core}$ &  Notes    \\
name       & Id.$^\oplus$  &         &  hh:mm:ss.ss        & dd:mm:ss.ss          &  hh:mm:ss.ss    &  dd:mm:ss.ss     &           \\
(1)        & (2)  &  (3)    &    (4)              &   (5)               &  (6)            & (7)              &   (8)   \\
\hline
\endhead
  \hline
\endfoot
  \hline
\endlastfoot
J0012$-$1050 & G  & (0.322) &  00:12:08.59        &$-$10:50:43.68       & 00:12:08.64     & $-$10:50:43.11   &  1        \\
J0040$-$1001& G   & (0.646) &  00:40:31.07        &$-$10:01:29.19       &                 &                  &  1      \\
J0142$-$0000& G   & (0.478) &  01:42:48.84        &$-$00:00:34.84       &                 &                  &  1     \\
J0202$+$0015&     &         &                     &                     &                 &                  &  1,2    \\
J0251$+$0039&     &         &                     &                     &                 &                  &  1,2    \\
J0848$+$2925&     &         &                     &                     &                 &                  &  1,2    \\
J0901$+$1212& G   & (0.473) &  09:01:31.55        &   $+$12:12:12.91    &                 &                  &  1      \\
J0927$+$2932& G   & (0.728) &  09:27:44.00        &   $+$29:32:34.43    &   09:27:43.88   &   $+$29:32:32.37    &  1,3    \\
J1009$+$3608&     &         &                     &                     &                 &                  &  2      \\
J1044$+$1442 & G  & 0.1547  &  10:44:34.63        &   $+$14:42:04.06    &                 &                  &  2       \\
J1057$+$5326&     &         &                     &                     &                 &                  &  2      \\
J1107$+$1216 & G  &         &                     &                     &                 &                  &  3       \\
J1114$+$3658&     &         &                     &                     &                 &                  &  2      \\
J1120$+$1611&     &         &                     &                     &                 &                  &  2      \\
J1123$+$4757& G   & (0.816) &  11:23:18.74        &   $+$47:57:59.26    &                 &                  &  1      \\
J1154$+$3021& S   &         &                     &                     &                 &                  &  2      \\
J1204$+$0756&     &         &                     &                     &                 &                  &  2      \\
J1233$-$0412&     &         &                     &                     &                 &                  &  2      \\
J1246$+$3848& G   & (0.814) &  12:46:08.23        &   $+$38:48:37.42    &                 &                  &  1      \\
J1248$+$2022&     &         &                     &                     &                 &                  &  1,2    \\
J1306$+$1922& G   & (0.573) &  13:06:00.63        &   $+$19:22:48.74    &                 &                  &  1      \\
J1308$+$3955& G   & (0.104) &  13:08:10.19        &   $+$39:55:05.42    &                 &                  &  1      \\
J1327$+$4946&     &         &                     &                     &                 &                  &  1,2    \\
J1332$+$1821& G   & 0.5049  &  13:32:24.76        &   $+$18:21:00.49    &                 &                  &  1      \\
J1335$+$2042&     &         &                     &                     &                 &                  &  2      \\
J1339$+$3756& G   & 0.3750  &  13:39:47.28        &   $+$37:56:52.82    & 13:39:47.26     & $+$37:56:52.10      &  1      \\
J1342$+$0225&     &         &                     &                     &                 &                  &  1,2    \\
J1342$+$0642& G   & (0.052) &  13:42:51.27        &   $+$06:42:31.05    &                 &                  &  1      \\
J1343$+$6306& S   &         &                     &                     &                 &                  &  2      \\
J1351$+$2519&     &         &                     &                     &                 &                  &  1,2      \\
J1352$-$0325&     &         &                     &                     &                 &                  &  2      \\
J1355$+$3525& G   & 0.1078  &  13:55:26.19        &   $+$35:25:44.12    &                 &                  &  1      \\
J1357$+$3407&     &         &                     &                     &                 &                  &  2      \\
J1404$+$5622&     &         &                     &                     &                 &                  &  2      \\
J1417$+$2508&     &         &                     &                     &                 &                  &  2      \\
J1428$+$2410&     &         &                     &                     &                 &                  &  2      \\
J1432$-$0507&     &         &  14:32:22.27        &   $-$05:07:15.49    &                 &                  &  1      \\
J1432$+$2634& G   & (0.648) &  14:32:40.76        &   $+$26:34:53.37    &                 &                  &  1      \\
J1434$+$2046& S   &         &  14:34:13.70        &   $+$20:46:13.74    & 14:34:13.57     & $+$20:46:11.44      &  1      \\
J1446$+$0047& S   &         &  14:46:36.92        &   $+$00:47:04.67    &                 &                  &  1      \\
J1448$+$0357&     &         &                     &                     &                 &                  &  2      \\
J1450$-$0607&     &         &                     &                     &                 &                  &  2      \\
J1459$+$3836&     &         &                     &                     &                 &                  &  1,2      \\
J1503$+$1154& S   &         &  15:03:11.75        &   $+$11:54:40.83    &                 &                  &  3      \\
J1511$+$5650& G   & 0.6319  &  15:11:09.21        &   $+$56:50:52.08    &  15:11:09.19    &  $+$56:50:51.71     &  1      \\
J1511$+$3633& G   & 0.1638  &  15:11:25.91        &   $+$36:33:36.03    &                 &                  &  1      \\
J1540$+$3121& S   &         &  15:40:13.86        &   $+$31:21:25.08    &                 &                  &  1      \\
J1540$+$6344&     &         &                     &                     &                 &                  &  2      \\
J1546$+$0844 & G  & 0.1849  &  15:46:35.78        & $+$08:44:24.63      &                 &                  &  1       \\
J1551$+$2654&     &         &                     &                     &                 &                  &  2      \\
J1554$+$3233 &    &         &                     &                     &                 &                  &  2      \\
J1604$+$2703&     &         &                     &                     &                 &                  &  1,2    \\
J1607$+$0625& G   & (0.612) &  16:07:19.46        &   $+$06:25:46.90    &                 &                  &  1      \\
J1611$+$1017 & G  & (0.737) &  16:11:14.84        & $+$10:17:12.81      &                 &                  &  1       \\
J1613$+$2929&     &         &                     &                     &                 &                  &  1,2  \\
J1613$+$5203& S   &         &  16:13:21.11        &  $+$52:03:01.75     &                 &                  &  1      \\
J1619$+$2841&     &         &                     &                     &                 &                  &  1,2      \\
J1642$+$4501 & G  &  0.1862 &  16:42:22.78        & $+$45:01:04.21      &                 &                  &   1      \\
J1706$+$3816&     &         &                     &                     &                 &                  &   2      \\
J1716$+$5023&     &         &                     &                     &                 &                  &   1,2   \\
J1720$+$3015& G   & (0.634) &  17:20:50.99        &   $+$30:15:39.29    &                 &                  &   1     \\
J1727$+$3727&     &         &                     &                     &                 &                  &   1,2  \\
J1733$+$4342&     &         &                     &                     &                 &                  &   1,2   \\
\end{longtable}
\footnotesize{Notes 1: higher-resolution images required to clarify the structure;
         2: optical identification is either unavailable or uncertain;
         3: separation of optical position from possible core $\sim$2 arcsec.\\
         $^\oplus$ In the second column G and S represent a galaxy and star respectively, as listed in SDSS.  }  
\end{center}
\egroup
}

\clearpage

{\small
\bgroup
%
%
%
\begin{center}
\begin{longtable}{cccccccc}

\caption{Sources classified as non-DDRGs.} \label{tab:sample_nonddrgs} \\
   \hline
Source     & Opt. &  $z$    &  RA$_{\rm optical}$ & Dec$_{\rm optical}$ & RA$_{\rm core}$ & Dec$_{\rm core}$ &  Notes    \\
name       & Id.$^\oplus$  &         &  hh:mm:ss.ss        & dd:mm:ss.ss          &  hh:mm:ss.ss    &  dd:mm:ss.ss     &           \\
(1)        & (2)  &  (3)    &    (4)              &   (5)               &  (6)            & (7)              &    (8)   \\ \hline
   \endfirsthead

{{ \tablename\ \thetable{} --(continued)}} \\
  \hline
Source     & Opt. &  $z$    &  RA$_{\rm optical}$ & Dec$_{\rm optical}$ & RA$_{\rm core}$ & Dec$_{\rm core}$ &  Notes    \\
name       & Id.$^\oplus$  &         &  hh:mm:ss.ss        & dd:mm:ss.ss          &  hh:mm:ss.ss    &  dd:mm:ss.ss     &           \\
(1)        & (2)  &  (3)    &    (4)              &   (5)               &  (6)            & (7)              &    (8)   \\ \hline
\endhead

  \hline
\endfoot

  \hline
\endlastfoot
J0013$-$0919& G   & (0.477) &  00:13:57.23        & $-$09:19:49.29      &                 &                  &         \\
J0032$-$0019& G   & (0.625) &  00:32:47.93        & $-$00:19:36.96      &                 &                  &           \\
J0056$-$1051& G   & 0.1959  &  00:56:41.46        & $-$10:52:03.11      & 00:56:41.46     & $-$10:52:03.44   &  1      \\
J0118$+$0114&     &         &                     &                     &                 &                  &         \\
J0729$+$3226& G   & (0.423) &  07:29:06.37        & $+$32:26:40.39      &                 &                  &         \\  
J0758$+$1617& G   & (0.626) &  07:58:30.23        & $+$16:17:36.26      &                 &                  &  3      \\
J0759$+$4051& G   & (0.602) &  07:59:12.22        & $+$40:51:54.79      & 07:59:12.22     & $+$40:51:54.42      &  1      \\    
J0838$+$2404& S   &         &  08:38:18.39        & $+$24:04:49.65      & 08:38:18.40     & $+$24:04:49.26      &  1      \\
J0902$+$5707& S   & 1.5963  &  09:02:07.20        & $+$57:07:38.09      & 09:02:07.22     & $+$57:07:37.95      &  1      \\
J0911$+$1255& G   & 0.0495  &  09:11:34.75        & $+$12:55:38.12      &                 &                  &         \\
J0912$+$0810& G   & (0.736) &  09:12:20.19        & $+$08:10:43.24      &                 &                  &         \\
J0914$+$1006&     &         &                     &                     &                 &                  &  4      \\ 
J0928$-$0319&     &         &  09:28:41.93        & $-$03:19:50.27      & 09:28:41.91     & $-$03:19:51.04   &  1      \\
J0942$+$2710& S   &         &  09:42:20.04        & $+$27:10:31.72	& 09:42:20.04     & $+$27:10:31.58      &  1      \\
J0948$+$5758& G   & (0.337) &  09:48:07.56        & $+$57:58:53.24      & 09:48:07.69     & $+$57:58:53.86      &  1      \\
J0949$+$2044& G   & (0.345) &  09:49:12.21        & $+$20:44:14.50      &                 &                  &         \\
J0953$+$1403& G   & 0.2376  &  09:53:42.24        & $+$14:03:58.01      &                 &                  &  1      \\
J0959$+$1558& G   & (0.767) &  09:59:08.01        & $+$15:58:30.32      &                 &                  &         \\
J1026$+$3639& G   & (0.816) &  10:26:07.98        & $+$36:40:01.21      & 10:26:07.99     & $+$36:40:01.47      &  1      \\
J1028$+$4306& S   &         &  10:28:33.11        & $+$43:06:26.93      & 10:28:33.14     &$+$ 43:06:26.94      &  1      \\
J1033$+$0755& S   &         &  10:33:40.07        & $+$07:55:57.7       & 10:33:40.06     & $+$07:55:58.10      &  1      \\ 
J1041$+$5233&     &         &                     &                     &                 &                  &  5      \\
J1049$+$0059& G   & 0.1064  &  10:49:14.08        & $+$00:59:45.26      & 10:49:14.07     & $+$00:59:45.18      &  1      \\
J1053$+$3125& G   & (0.647) &  10:53:04.71        & $+$31:26:01.34      & 10:53:04.70     & $+$31:26:01.13      &  1      \\
J1054$+$0740& G   & 0.0968  &  10:54:52.10        & $+$07:40:06.37      & 10:54:52.08     & $+$07:40:06.74      &  1      \\
J1103$+$0249& G   & (0.316) &  11:03:38.13        & $+$02:49:28.81      &                 &                  &         \\
J1105$+$2317&     &         &                     &                     &                 &                  &         \\
J1139$+$6030& G   & (0.707) &  11:39:31.52        & $+$60:33:15.18      &                 &                  &         \\   
J1141$+$0802& G   & 0.2282  &  11:41:25.98        & $+$08:02:16.51      & 11:41:26.01     & $+$08:02:16.90      &  1      \\
J1150$+$4046& G   & (0.380) &  11:50:55.10        & $+$40:46:32.09      & 11:50:55.11     & $+$40:46:31.31      &  1      \\
J1201$+$2256& G   & 0.2594  &  12:01:41.70        & $+$22:56:46.41      & 12:01:41.68     & $+$22:56:46.05      &  1      \\
J1206$+$1626& G   & 0.3036  &  12:06:03.54        & $+$16:26:34.92      &                 &                  &  2      \\ 
J1208$+$2220& G   & (0.636) &  12:08:21.81        & $+$22:20:03.89      & 12:08:21.99     & $+$22:19:58.37      &         \\
J1213$+$1343& G   & 0.1743  &  12:13:06.68        & $+$13:43:17.79      &                 &                  &         \\
J1214$+$5107& G   & (0.718) &  12:14:46.64        & $+$51:07:04.30      &                 &                  &         \\    
J1224$+$0203& S   & 0.4492  &  12:24:25.61        & $+$02:03:09.68      & 12:24:25.54     & $+$02:03:10.82      &  1      \\
J1224$+$2358& G   & (0.612) &  12:24:39.31        & $+$23:58:56.69      &                 &                  &         \\  
J1230$+$1046&     &         &                     &                     &                 &                  &         \\
J1234$+$5753& G   & 0.1529  &  12:34:24.35        & $+$57:53:27.33      & 12:34:24.26     & $+$57:53:26.08      &         \\
J1240$+$5334& G   & (0.267) &  12:40:12.46        & $+$53:34:37.37      & 12:40:12.49     & $+$53:34:37.55      &         \\
J1242$+$4244& G   & (0.388) &  12:42:37.95        & $+$42:44:03.09      &                 &                  &         \\
J1243$+$1508& S   &         &  12:43:44.83        & $+$15:08:20.66      & 12:43:44.83     & $+$15:08:20.61      &  1      \\
J1248$+$1725& G   & (0.581) &  12:48:04.08        & $+$17:25:50.56      &                 &                  &  6      \\
J1248$-$0301& Q   & 1.0337  &  12:48:04.09        & $-$03:01:14.54      & 12:48:04.11    &$-$03:01:13.27    &  1      \\
J1248$+$5942& G   & (0.557) &  12:48:35.68        & $+$59:42:22.37      &                 &                  &         \\
J1255$+$4405& G   & (0.211) &  12:55:54.59        & $+$44:05:21.82      &                 &                  &  6      \\
J1319$+$0502& S   &         &  13:19:43.58        & $+$05:02:43.02      & 13:19:43.54     & $+$05:02:43.02      &  1      \\ 
J1325$+$5736& G   & (0.139) &  13:25:11.17        & $+$57:36:01.58      &                 &                  &  2      \\
J1339$+$2812& S   &         &  13:39:04.29        & $+$28:12:41.19      & 13:39:04.29     & $+$28:12:41.18      &  1      \\
J1339$-$0637&     &         &  13:39:07.10        & $-$06:37:04.96      & 13:39:07.22     & $-$06:37:05.09   &  1      \\
J1339$+$1024& G   & (0.464) &  13:39:13.64        & $+$10:24:47.88      & 13:39:13.42     & $+$10:24:46.44      &  1      \\
J1343$+$4627& G   & 0.2248  &  13:43:00.36        & $+$46:27:19.99	& 13:43:00.36     & $+$46:27:19.87      &  1      \\
J1344$+$3317& Q   & 0.6862  &  13:44:15.75        & $+$33:17:19.13      & 13:44:15.72     & $+$33:17:18.74      &  1      \\
J1349$-$0149& G   & 0.2098  &  13:49:18.71        & $-$01:49:21.40      & 13:49:18.76     & $-$01:49:20.22   &  2      \\
J1351$+$0728& G   & 0.1500  &  13:51:10.81        & $+$07:28:46.07      & 13:51:10.81     & $+$07:28:46.78      &  7      \\
J1400$+$0736& G   & (0.718) &  14:00:06.93        & $+$07:37:00.48      &                 &                  &         \\    
J1402$+$6105& G   & (0.735) &  14:02:14.73        & $+$61:05:31.35      &                 &                  &         \\
J1410$+$3749&     &         &                     &                     &                 &                  &  8      \\
J1412$+$2301& G   & (0.486) &  14:12:50.38        & $+$23:01:16.61      &                 &                  &         \\
J1413$+$0741& G   & (0.539) &  14:13:15.25        & $+$07:41:43.63      &                 &                  &         \\
J1420$+$3552& G   & (0.549) &  14:20:37.55        & $+$35:52:51.18      &                 &                  &         \\
J1423$+$2448& G   & (0.585) &  14:23:21.33        & $+$24:48:15.05      & 14:23:21.24     & $+$24:48:15.07      &  1      \\
J1425$-$0456&     &         &  14:25:12.21        &$-$04:56:34.71       & 14:25:12.25     &$-$04:56:35.70    &  2      \\
J1430$+$5217& G   & 0.3675  &  14:30:17.34        & $+$52:17:35.32      & 14:30:17.14     & $+$52:17:35.74      &  6      \\
J1431$+$0538& G   & (0.245) &  14:31:03.50        & $+$05:38:12.45	&                 &                  &         \\
J1431$+$1922& G   & 0.2136  &  14:31:49.14        & $+$19:23:00.12      & 14:31:49.15     & $+$19:23:59.83      &  1      \\
J1434$+$0441& S   &         &  14:34:21.56        & $+$04:41:37.21      & 14:34:21.57     & $+$04 41 38.23      &  1      \\
J1437$+$2445& G   & 0.0862  &  14:37:15.00        & $+$24:45:32.21      & 14:37:15.02     & $+$24:45:33.21      &  1      \\
J1439$+$5314& G   & (0.266) &  14:39:34.48        & $+$53:14:37.44      & 14:39:34.47     & $+$53:14:39.65      &  1      \\
J1439$+$2824& G   & (0.367) &  14:39:58.41        & $+$28:24:22.66	& 14:39:58.44     & $+$28:24:22.89      &  1      \\
J1444$+$3817&     &         &                     &                     &                 &                  &         \\
J1445$-$0626&     &         &                     &                     &                 &                  &         \\
J1448$+$2954&     &         &                     &                     &                 &                  &  9      \\  
J1448$+$4923& G   & (0.494) &  14:48:59.59        & $+$49:23:40.84      &                 &                  &         \\
J1450$+$0001& Q   & 1.9679  &  14:50:49.93        & $+$00:01:44.34      & 14:50:49.94     & $+$00:01:44.25      &         \\
J1450$-$0315&     &         &  14:50:54.99        & $-$03:15:52.32      &                 &                  &         \\
J1459$+$1655& G   & (0.608) &  14:59:36.39        & $+$16:55:25.59      & 14:59:36.43     & $+$16:55:25.62      &  1      \\
J1500$+$1327& G   & 0.1116  &  15:00:03.96        & $+$13:27:45.57      & 15:00:03.99     & $+$13:27:45.80      &  1,2    \\
J1501$+$5455& G   & 0.3388  &  15:01:17.98        & $+$54:55:18.37      & 15:01:17.95     & $+$54:55:18.02      &  1      \\
J1501$+$4012& G   & (0.466) &  15:01:21.53        & $+$40:12:19.88      & 15:01:21.57    & $+$40:12:20.04      &  1      \\ 
J1525$+$1253& G   & 0.2571  &  15:25:11.68        & $+$12:52:58.03      & 15:25:11.68     & $+$12:52:57.83      &  1      \\
J1527$+$1822& G   & (0.472) &  15:27:37.11        & $+$18:22:51.31      & 15:27:37.10     & $+$18:22:50.80      &  1      \\
J1530$+$2316& G   & 0.0899  &  15:30:07.96        & $+$23:16:16.02      & 15:30:07.94     & $+$23:16:15.74      &  1      \\
J1530$-$0703&     &         &  15:30:58.81        & $-$07:03:31.74      & 15:30:58.90     & $-$07:03:32.40   &  1      \\
J1540$+$4925& G   & (0.472) &  15:40:28.65        & $+$49:25:14.72      & 15:40:28.67     & $+$49:25:14.55     &  1      \\
J1540$+$0132& Q   & 0.7743  &  15:40:47.88        & $+$01:32:07.17      & 15:40:47.88     & $+$01 32 06.84      &  1      \\ 
J1545$-$0330&     &         &  15:45:37.96        & $-$03:30:46.63      &                 &                  &         \\ 
J1548$+$3633& G   & 0.1963  &  15:48:05.70        & $+$36:33:40.57      & 15:48:05.74     & $+$36:33:40.72      &  1      \\
J1550$+$2246& G   & (0.435) &  15:50:05.00        & $+$22:46:03.46      & 15:50:04.97     & $+$22 46 05.82      &  1      \\
J1557$+$1618& G   & 0.0370  &  15:57:49.61        & $+$16:18:36.59      &                 &                  &         \\
J1558$+$0759& G   & (0.227) &  15:58:34.46        & $+$07:59:46.13      &                 &                  &  2      \\
J1600$+$1306& S   &         &  16:00:58.77        & $+$13:06:59.59      & 16:00:58.79     & $+$13:06:59.64      &  1      \\
J1601$+$3423& G   & (0.669) &  16:01:30.46        &  $+$ 34:23:12.42    &                 &                  &         \\
J1608$+$2828& G   & 0.0502  &  16:08:21.14        & $+$28:28:43.29      & 16:08:21.15     & $+$28:28:43.54      &  1,2    \\
J1614$+$3210& G   & (0.663) &  16:14:49.71        & $+$32:10:54.46      &                 &                  &         \\
J1614$+$0240&     &         &                     &                     &                 &                  &  2      \\
J1615$+$1245&     &         &                     &                     &                 &                  &         \\
J1616$+$2809&     &         &                     &                     &                 &                  &         \\
J1617$+$5451& S   &         &  16:17:56.89        & $+$54:51:13.80      & 16:17:56.94     & $+$54:51:13.90      &  1,2    \\
J1618$+$1211& G   & 0.3934  &  16:18:03.56        & $+$12:11:32.62	& 16:18:03.60     & $+$12:11:33.70      &  1      \\
J1620$+$2017& G   & (0.056) &  16:20:21.74        & $+$20:17:07.32      &                 &                  &  1      \\
J1623$+$4318& G   & (0.687) &  16:24:00.24        & $+$43:18:41.14      & 16:24:00.27     & $+$43:18:41.65      &  1      \\
J1627$+$5019& G   & (0.450) &  16:27:24.54        & $+$50:19:40.92      & 16:27:24.54     & $+$50:19:40.89      &  1      \\
J1628$+$3906& G   & (0.499) &  16:28:11.44        & $+$39:06:36.70      & 16:28:11.42     & $+$39:06:36.65      &  1      \\
J1628$+$5750& G   & (0.262) &  16:28:57.46        & $+$57:50:44.77	&                 &                  &         \\
J1631$+$1855& G   & (0.107) &  16:31:04.91        & $+$18:55:22.77	& 16:31:04.95     & $+$18:55:22.26      &  1      \\
J1633$+$0847& G   & (0.191) &  16:33:00.85        & $+$08:47:36.44      & 16:33:00.83     & $+$08:47:36.61      &  1,7    \\ 
J1637$+$4130& S   & 1.1781  &  16:37:02.20        & $+$41:30:22.22      &                 &                  &         \\
J1643$+$2642&     &         &                     &                     &                 &                  &  6      \\
J1646$+$5549&     &         &  16:46:19.05        & $+$55:49:58.13      & 16:46:19.19     & $+$55:49:58.54      &  1      \\
J1647$+$5154& G   & (0.303) &  16:47:05.21        & $+$51:54:11.47      &                 &                  &  2      \\
J1649$+$1651& G   & (0.563) &  16:49:48.74        & $+$16:51:47.14      &                 &                  &         \\
J1658$+$3408&     &         &                     &                     &                 &                  &         \\
J1659$+$2602& G   & (0.494) &  16:59:52.81        & $+$26:02:38.69      & 16:59:52.80     & $+$26:02:39.42      &         \\
J1700$+$4851& G   & (0.204) &  17:00:35.32        & $+$48:51:03.95      &                 &                  &         \\
J1702$+$5312&     &         &  17:02:01.36        & $+$53:12:19.02      & 17:02:01.38     & $+$53:12:34.37      &  1      \\            
J1703$+$5533&     &         &                     &                     &                 &                  &         \\
J1705$+$2839& G   & (0.300) &  17:05:02.05        & $+$28:39:06.94      &                 &                  &         \\
J1710$+$4611& G   & (0.759) &  17:10:36.66        &   $+$46:11:11.75    &                 &                  &         \\
J1710$+$4239& G   & (0.176) &  17:10:40.73        & $+$42:39:45.08      & 17:10:40.75     & $+$42:39:44.80      &   1     \\
J1713$+$5812&     &         &                     &                     &                 &                  &         \\
J1723$+$3017& G   & (0.645) &  17:23:08.07        & $+$30:17:16.32      & 17:23:08.04     & $+$30:17:15.79      &   6     \\
J1725$+$3452& S   &         &  17:25:07.68        & $+$34:52:15.97      &                 &                  &         \\
J1728$+$4855&     &         &                     &                     &                 &                  &   9     \\
J1728$+$4421& G   & (0.569) &  17:28:56.98        & $+$44:21:37.30      &                 &                  &         \\ 
J1732$+$5634& G   &  0.3330 &  17:32:50.22        & $+$56:34:26.61      & 17:32:50.18     & $+$56:34:27.07      &  1,2    \\
J2143$-$0858&     &         &                     &                     &                 &                  &         \\
J2149$-$0004& G   & (0.650) &  21:49:01.07        & $-$00:04:22.19      &                 &                  &         \\
J2259$-$0925&     &         &                     &                     &                 &                  &         \\
J2315$-$0026& G   & 0.0909  &  23:15:42.11        & $-$00:26:07.05      & 23:15:42.09     & $-$00:26:07.34   &  1      \\
J2322$-$0941& G   & (0.273) &  23:22:08.21        & $-$09:41:58.00      & 23:22:08.25     & $-$09 41 58.67   &  1      \\
J2353$+$0012& G   & 0.1663  &  23:53:55.59        & $+$00:12:56.40      & 23:53:55.65     & $+$00:12:56.12      &         \\   
\end{longtable}
\footnotesize{Notes 1: optical identification is associated with one of the inner objects;
2: FRI type structure;
3: although inner lobe appears larger, high-resolution observations would be helpful to confirm its classification;
4: optical identification unclear (there are two possible candidates);
5: northern compact component possibly unrelated;
6: X-shaped source;
7: two independent sources nearby;
8: optical object associated with westernmost component;
9: two optical objects along the radio axis with the western one being the likely identification;
10: optical object associated with the northernmost component.\\
$^\oplus$ In the second column G, S and Q  represent  galaxy, star and quasi stellar object (QSO), as listed
in SDSS.}  
\end{center}
\egroup

}

\clearpage
\begin{table}
\caption{Candidate DDRGs from the FIRST survey classified as WATs.}
\label{tab:sample_WAT}
{\small
\centering
\begin{tabular}{c c c c c c c }
\hline
Source     & Opt. &  $z$    &  RA$_{\rm optical}$ & Dec$_{\rm optical}$ & RA$_{\rm core}$ & Dec$_{\rm core}$   \\
name       & Id.$^\oplus$  &         &  hh:mm:ss.ss        & dd:mm:ss.ss          &  hh:mm:ss.ss    &  dd:mm:ss.ss       \\
(1)        & (2)  &  (3)    &    (4)              &   (5)               &  (6)            & (7)                \\
\hline
J0041$-$0925 & G  & 0.05779 &  00:41:50.17        & $-$09:25:47.43      & 00:41:50.12     & $-$09:25:47.99      \\
J0123$-$1025& G   & 0.1537  &  01:23:12.52        & $-$10:25:10.38      &                 &                    \\
J0930$+$5930& G   & (0.618) &  09:30:06.88        & $+$59:30:32.82      & 09:30:06.86     & $+$59:30:32.44        \\
J0948$+$4218& G   & (0.372) &  09:48:42.95        & $+$42:18:31.35      & 09:48:42.98     & $+$42:18:31.33        \\
J1028$+$0345 & G  & (0.108) &  10:28:23.47        & $+$03:45:31.51      &                 &                     \\
J1035$+$4255 & G  & 0.13602 &  10:35:02.61        & $+$42:55:48.34      & 10:35:02.59     & $+$42 55 47.68         \\
J1046$+$1210 &    &         &                     &                     &                 &                     \\
J1119$+$5808 &    &         &                     &                     &                 &                     \\
J1151$+$0422 & G  & (0.150) &  11:51:46.90        & $+$04:22:22.73      &                 &                     \\
J1153$+$2341& S   &         &  11:53:18.07        & $+$23:41:13.52      & 11:53:18.15     & $+$23:41:13.22         \\
J1258$+$1110 & G  & (0.699) &  12:58:22.78        & $+$11:10:39.56      &                 &                     \\
J1335$+$5352 & S  &         &  13:35:40.85        & $+$53:52:24.40      &                 &                     \\
J1342$+$0643& G   & (0.619) &  13:42:38.95        & $+$06:43:25.33      &                 &                     \\
J1406$+$5504& G   &  0.2505 &  14:06:55.11        & $+$55:04:02.87      & 14:06:54.72     & $+$55 04 01.72         \\
J1421$-$0743 &    &         &                     &                     &                 &                     \\
J1432$+$4436& G   & 0.1815  &  14:32:43.85        & $+$44:36:14.21      & 14:32:43.87     & $+$44:36:14.75         \\
J1453$+$2108& G   & (0.634) &  14:53:05.17        & $+$21:08:31.64      & 14:53:05.18     & $+$21:08:31.50         \\
J1458$+$2754 & G  & 0.2288  &  14:59:00.23        & $+$27:54:00.78      & 14:59:00.20     & $+$27:54:00.10         \\
J1510$+$0544 & G  & (0.146) &  15:10:56.10        & $+$05:44:41.19      &                 &                     \\
J1534$+$0556 & G  & (0.257) &  15:34:39.92        & $+$05:56:17.45      &                 &                     \\
J1547$-$0046 &    &         &                     &                     &                 &                     \\
J1706$+$2006 & G  & (0.305) &  17:06:16.33        & $+$20:06:48.98      &                 &                     \\
J2151$-$0005 & G  & (0.810) &  21:51:27.90        & $-$00:05:07.87      &                 &                     \\
\hline
\end{tabular}

\footnotesize
$^\oplus$ In the second column G and S represent galaxy and star respectively, as listed in SDSS.

}
\end{table}

\label{lastpage}
\end{document}